\documentclass[sigconf,natbib=true,anonymous=false]{acmart}

\usepackage{graphicx} 
\usepackage{colortbl}
\usepackage{flafter} 

\newcommand{\specialcellleft}[2][l]{%
  \begin{tabular}[#1]{@{}l@{}}#2\end{tabular}}

\title{Enhancing Travel Decision-Making: A Contrastive Learning Approach for Personalized Review Rankings in Accommodations}

\author{Reda Igebaria}
\authornote{Both authors contributed equally to this research.}
\email{redaigbaria@campus.technion.ac.il}
\affiliation{\institution{Booking.com, Tel Aviv} \country{Israel}}

\author{Eran Fainman}
\authornotemark[1]
\email{eran.fainman@booking.com}
\affiliation{\institution{Booking.com, Tel Aviv} \country{Israel}}

\author{Sarai Mizrachi}
\email{sarai.mizrachi@booking.com}
\affiliation{\institution{Booking.com, Tel Aviv} \country{Israel}}

\author{Moran Beladev}
\email{moran.beladev@booking.com}
\affiliation{\institution{Booking.com, Tel Aviv} \country{Israel}}

\author{Fengjun Wang}
\email{fengjun.wang@booking.com}
\affiliation{\institution{Booking.com, Amsterdam} \country{Netherlands}}

\date{January 2024}

\begin{document}

\begin{abstract}
User-generated reviews significantly influence consumer decisions, particularly in the travel domain when selecting accommodations. This paper contribution comprising two main elements. Firstly, we present a novel dataset of authentic guest reviews sourced from a prominent online travel platform, totaling over two million reviews from 50,000 distinct accommodations. Secondly, we propose an innovative approach for personalized review ranking. Our method employs contrastive learning to intricately capture the relationship between a review and the contextual information of its respective reviewer. Through a comprehensive experimental study, we demonstrate that our approach surpasses several baselines across all reported metrics. Augmented by a comparative analysis, we showcase the efficacy of our method in elevating personalized review ranking. The implications of our research extend beyond the travel domain, with potential applications in other sectors where personalized review ranking is paramount, such as online e-commerce platforms.

\end{abstract}

\maketitle
\section{Introduction}
Large online travel platforms allow millions of users to book stays worldwide. The influence of user-generated reviews has become pivotal in the decision-making process for those seeking to book accommodations \cite{gretzel2008use,baka2016becoming,tuominen2011influence,ricci2006product}. The growing volume of reviews underscores the imperative for an effective review ranking mechanism. The one-size-fits-all approach of traditional review ranking models poses a significant challenge in catering to the diverse preferences and priorities of individual users. Current algorithms often prioritize reviews based on helpfulness votes \cite{10.1145/3437120.3437308,10.1145/1282100.1282158,KORFIATIS2012205,yang-etal-2015-semantic,5590249}, introducing biases and neglecting the nuanced perspectives of users with distinct needs \cite{yue2010beyond,agarwal2018consistent,wang2016learning}. One common issue is the sparsity of helpfulness votes, with most reviews lacking any such votes \cite{kuan2015makes}. In our dataset, this problem is compounded by the fact that only 8.7\% of reviews receive helpful votes, making the signal notably sparse. Furthermore, these votes are anonymous which makes it not feasible to use them for developing personalized review ranking models.

In response to these limitations, we present a novel approach for personalized reviews ranking. Our approach acknowledges that different users have different characteristics (like families, solo travelers, couples etc), different trip types (e.g., beach, city, nature, etc) and therefore prioritize different accommodation features (like accessibility, sustainability, inclusivity etc). We consider these user and accommodation characteristics as context. We leverage contrastive learning \cite{bengio2013representation} to capture the intricate relationship between users' context and their reviews. This approach relies on that user-generated reviews reflect their personal experience with a focus on what they most liked and disliked about their stay, taking into account their personal preferences and the key elements that were crucial to their stay \cite{baka2016becoming}.

Publishing a dataset of user-generated reviews holds paramount importance in the travel domain due to its pivotal role in facilitating informed decision-making for users who book accommodations. Unlike e-commerce, where review datasets are relatively abundant, the availability of comprehensive accommodation review datasets remains scarce. This scarcity poses a significant challenge for researchers, developers, and industry stakeholders seeking to enhance tourism products and services. Events like the Rectour workshop, held annually within the RecSys conference, underscore the growing recognition of the critical role that tourism review datasets play in advancing recommendation systems tailored specifically to the complexities of the tourism industry. Thus, while e-commerce benefits from an abundance of review data, the tourism sector faces a pressing need for the availability of comprehensive datasets to drive innovation and enhance tourism experiences.

As a part of this work, we publish a dataset that consists of over two million reviews from 50,000 unique accommodations. It contains contextual details about the guest who wrote the review (e.g., number of booked nights and guest type), the accommodation (e.g., accommodation type and average review score), and the review itself (including its textual fields, number of helpful votes and the overall rating score). The process of curating the dataset leveraged the Text2topic model \cite{wang2023text2topic} to select informative reviews, i.e., reviews that provide insights about topics related to the stay (such as cleanliness, value for money, breakfast etc).

In the following sections, we describe the details of our dataset curation and our modeling approach, present the findings of our experiments, and showcase the effectiveness of our methodology throughout a comparative analysis.

This paper contributions are summarized as follows:
\begin{itemize}
    \item We present a new large-scale, real-world dataset of informative reviews sourced from a major online travel platform. This dataset is publicly accessible via GitHub\footnote{\url{https://github.com/bookingcom/ml-dataset-reviews}}.
    \item We introduce a novel formulation for the personalized ranking task, which is designed to rank reviews based on their relevance to the reviewer's context.
    \item We propose a contrastive learning approach to tackle the task, utilizing a unique in-accommodation batch sampling method.
    \item Experimental results, conducted across various settings, consistently demonstrate that our method significantly outperforms ranking based on helpful votes. This is evident in terms of Mean Reciprocal Rank (MRR), precision@1, and precision@10 metrics.
    \item We provide interpretability and explainability of our model outputs by performing a comparative analysis highlighting common topics between reviews ranked by our model and reviews written by the users with the given context.
\end{itemize}

\section{Related Work}
In this section, we delve into the existing literature related to review ranking. We begin by providing an overview of both non-contextual and contextual review ranking methodologies, highlighting their key features and differences. Subsequently, we present an outline of the currently available datasets in this field, comparing and contrasting them with our dataset. Lastly, we examine contrastive learning, a technique that is leveraged by our work.

\subsection{Non-contextual Review Ranking}
\label{rw_supervised_helpfulness}
Early attempts on review helpfulness prediction were focused on predicting helpfulness using simple ML models such as SVM and linear regression \cite{10.1145/1282100.1282158}. These were done with hand-crafted features such as review length, readability \cite{KORFIATIS2012205}, sentiment \cite{yang-etal-2015-semantic} and subjectivity of the review \cite{5590249}. \citet{nayeem-rafiei-2023-role} used reviewer and temporal information in the decision making. Specifically, they leveraged mean votes of the user's past reviews and applied a time decay over the review age. The motivation for time decay is that old reviews might become irrelevant over time, e.g., a hotel might address cleaning issues by improving inspection following negative reviews. \citet{10.1371/journal.pone.0226902} empirically analyzed features that have been used in 149 previous papers. They found that semantic features (i.e., TF-IDF vector and pre-trained word embeddings) are the most predictive of helpfulness compared to other features related to sentiment, readbility, structure and syntax. \citet{RePEc:spr:elcore:v:23:y:2023:i:4:d:10.1007_s10660-022-09560-w} showed that fine-tuned BERT model outperforms bag-of-words approaches in identifying helpful reviews. \citet{10.1145/3308558.3313523} implemented a Bi-LSTM based model that predicts helpfulness given the review and product description. Finally, \citet{han-etal-2022-sancl, REN2024103573} used a multi-modal approach and incorporated images from the reviews into the model.

The studies cited above established ground truth labels based on the number of helpful votes. They approached the task of predicting helpfulness as a supervised binary classification problem, setting a threshold on the number of votes to categorize reviews as helpful or not. Notably, these studies did not incorporate user features or personalized mechanisms into their ranking methodologies, which presents a gap we aim to address in our model. Moreover, these methods exhibit a presentation bias, as users tend to vote on reviews they read, influenced by the original ranking \cite{yue2010beyond}. Our approach mitigates these biases by explicitly modeling the relationship between the review content and the corresponding reviewer's context, thereby removing the dependency on helpful votes.

\subsection{Context-aware Review Ranking}
Another line of work addresses the task of personalized reviews ranking. This task differs from the task discussed in the previous section in two key aspects: (1) the user's context is taken into account, and (2) the ground truth is user subjective, i.e., the same review might be helpful to one user but not to another. \citet{10.1145/2063576.2063938} employed traditional recommendation methods like matrix factorization to suggest reviews based on a latent representation of the user. \citet{10.1145/2513549.2513554} integrated social relations between users into the ranking, by constructing a social graph and integrating the connections as features for a linear regression model. Furthermore, \citet{peddireddy2020personalized} utilized a weighted term-frequency vector representing past user interactions, e.g. product purchases and reviews, and used Okapi BM25 \cite{robertson1995okapi} to rank the reviews. Lastly, \citet{10.1145/3414841} adopted a graph-based methodology to determine similarity between the user and reviewers, and rank accordingly.

While these studies aim to deliver more personalized reviews, they may encounter the cold-start problem when users lack sufficient interactions \cite{jalan2017context}. Additionally, they rely on non-anonymous helpful votes, which may not always be available and could be subject to presentation bias \cite{10.1145/2063576.2063938, 10.1145/2513549.2513554}. In contrast, our approach utilizes a context-aware strategy that does not rely on user-attributed helpful votes and is immune to the influence of review rank.

\subsection{User-generated Review Datasets}
There are several public user-generated review datasets in the travel domain, mainly crawled from leading online travel platforms such as TripAdvisor and Booking.com. For example, \citet{FANG2016498} crawled 41k reviews for attractions in New Orleans to analyze the perceived helpfulness of reviews (judged by the number of votes). \citet{10.1145/3437120.3437308} crawled 65k English hotel reviews from TripAdvisor and trained a DNN on helpfulness prediction. Finally, the largest dataset\footnote{\label{kaggle}\url{https://www.kaggle.com/datasets/jiashenliu/515k-hotel-reviews-data-in-europe/discussion}} we could find contains 515k reviews crawled from Booking.com published between August 2015 to August 2017.

In terms of scale, public review datasets in the travel domain contain hundreds of thousands of reviews at most. However, larger datasets exist in non-travel domains, e.g., Amazon Product Reviews dataset \cite{10.1145/2766462.2767755} contains 82.8 million reviews across 24 domains. Furthermore, the available contextual information about the user varies between the datasets. For example, the Booking.com crawled dataset\footref{kaggle} contains the reviewer's origin country and review publication date but doesn't include information about the stay (such as number of nights and traveler type).

The dataset we provide is a large-scale dataset with millions of user-generated accommodation reviews. It contains comprehensive information about the review and the reviewer's context. The dataset details and its curation process are described in Section \ref{sec:dataset}.

\subsection{Contrastive Learning}
Contrastive Learning (CL) is a powerful paradigm that effectively captures intricate relationships within data. Introduced as a technique to learn representations by contrasting positive pairs against negative pairs, CL aims to maximize the similarity between instances that should be similar while minimizing the similarity between those that should differ. This concept was first introduced by \citet{bengio2013representation}, and gained significant popularity and widespread attention with the introduction of the CLIP model \cite{radford2021learning}. In the CLIP framework, CL was shown to be effective when applied to a joint pre-training of visual and textual representations.

CL has since found applications beyond computer vision, including natural language processing and recommendation systems \cite{liu2021contrastive}. The versatility of CL lies in its ability to distill intricate patterns and semantic relationships from unlabeled data, significantly reducing the reliance on labeled datasets.

In the context of our study, we harness the potential of CL to model the nuanced relationship between reviewers' contexts and their corresponding reviews. Our approach contrasts positive instances, where reviews are paired with their reviewers' contexts, against negative instances, where reviews are paired with their non-reviewers' contexts from the same accommodation.

\section{Dataset}
\label{sec:dataset}
\begin{figure*}[ht]
    \centering
\includegraphics[width=0.88\textwidth,height=0.22\textwidth]{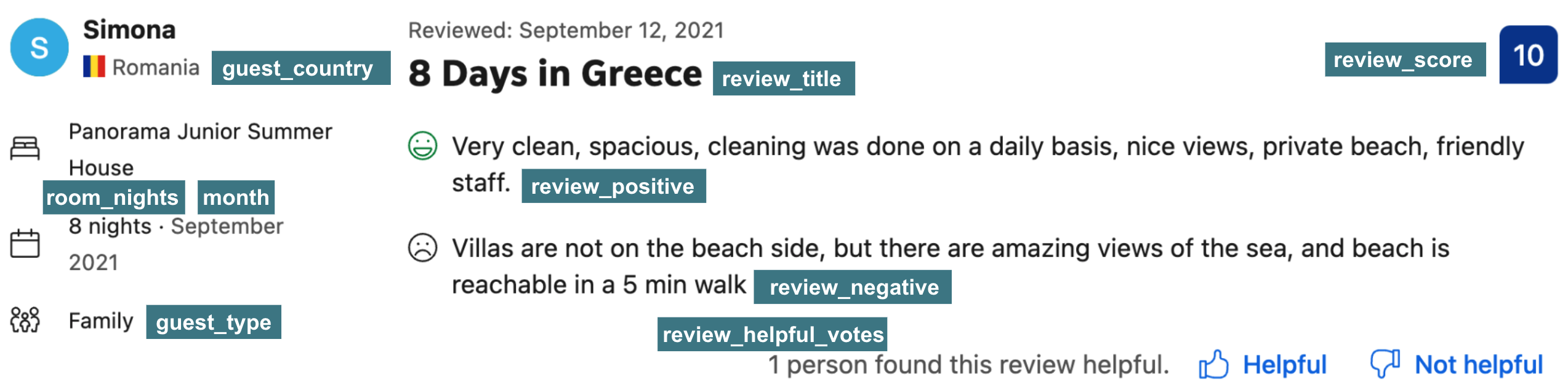}
    \caption{A guest review example. The respective names of the fields in our dataset are mentioned in green rectangles.}
    \label{fig:review_exp}
\end{figure*}

The dataset we publish contains authentic user-generated reviews from 50,000 accommodations. This includes information on the user reservation, the review and the accommodation. The dataset is from a leading online travel platform allowing reviews only from guests who stayed at the property. Therefore, every review is associated with the guest context, such as the number of nights, month, and the traveller type (e.g., solo traveller, family etc). There are several fields describing the review: the title, the positive ("liked") section, the negative ("disliked") section, the overall review score and the number of helpful votes. Figure~\ref{fig:review_exp} demonstrates an example of a guest review. The dataset is published under a non-commercial license\footnote{\url{https://creativecommons.org/licenses/by-sa/4.0/deed.en}}, available via GitHub\footnote{\url{https://github.com/bookingcom/ml-dataset-reviews}}.

\subsection{Data Selection}
The dataset consists of English reviews published in 2023. All reviews have passed a moderation process ensuring they are genuine and do not violate the platform guidelines\footnote{See more details in \url{https://www.booking.com/reviews_guidelines.html}}. In order to preserve user privacy, no personally identifiable information was included in the data. Similarly, to protect business-sensitive statistics, the dataset is limited to only tens of thousands of accommodations.

To identify informative reviews, we utilize a topic detection model called Text2topic, specifically designed for travel-related tasks by \citet{wang2023text2topic}. We select reviews that have at least three topics. 

Our observations \footnote{Based on the browser versions of the Booking.com and Expedia travel platforms} indicate that leading online travel platforms show 10 reviews in a page. Therefore, we select accommodations with at least 10 reviews. Finally, we uniformly sample 50,000 accommodations.

\begin{figure}[ht]
    \centering
\includegraphics[width=0.42\textwidth,height=0.315\textwidth]{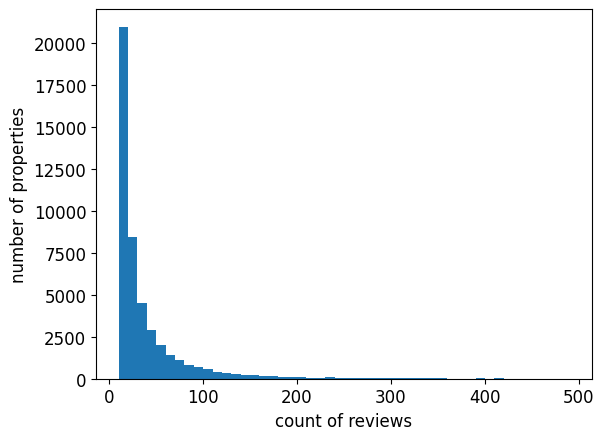}
        \caption{Histogram of number of reviews per accommodation}
    \label{fig:review_dist}
\end{figure}

Figure~\ref{fig:review_dist} shows the distribution of the number of reviews per accommodation. 36.8 is the average, 20 is the median, and 1,818 is the maximum.  

\subsection{Data Schema}
The dataset consists of 2,031,914 reviews, along with guest and accommodation context. Table~\ref{tab:dataset} describes the dataset fields. Key statistics of the data are described in table~\ref{tab:stat}.

\begin{table*}[!h]
	\centering
	\caption{Dataset description: review, user and accommodation fields}
	\label{tab:dataset}
	\begin{tabular}{p{0.25\textwidth}p{0.7\textwidth}}
		\toprule
\textbf{Field name} & \textbf{Description}\\
\midrule

\textbf{review\_title} & 
Review title\\
\hline
\textbf{review\_positive} & 
Positive ("liked") section in review\\
\hline
\textbf{review\_negative} &
Negative ("disliked") section in review \\
\hline
\textbf{review\_score} &
Overall review score for the stay\\
\hline
\textbf{review\_helpful\_votes} &
How many users marked the review as helpful\\
\hline
\textbf{guest\_type} &
\specialcellleft{
There are 4 traveller types: Solo traveller (1 adult) / Couple (2 adults) / \\Group (>2 adults) / Family with children (adults \& children)} \\
\hline
\textbf{guest\_country} &
\specialcellleft{Anonymized country from which the reservation was made}\\
\hline
\textbf{room\_nights} &
The length of the reservation, i.e. number of nights booked\\
\hline
\textbf{month} &
\specialcellleft{The month of the check-in date of the reservation}\\
\hline
\textbf{accommodation\_id} &
Anonymized accommodation ID \\
\hline
\textbf{accommodation\_type} &
\specialcellleft{The type of the accommodation, e.g. hotel, apartment, hostel}\\
\hline
\textbf{accommodation\_score} &
\specialcellleft{The overall average guest review score of the accommodation}\\
\hline
\textbf{accommodation\_country} &
Country of the accommodation\\
\hline
\textbf{accommodation\_star\_rating} &
\specialcellleft{Accommodation star rating is provided by the property, and is usually\\determined by an official accommodation rating organisation or another third party}\\
\hline
\textbf{location\_is\_beach} &
Is the accommodation located in a beach location\\
\hline
\textbf{location\_is\_ski} &
Is the accommodation located in a ski location\\
\hline
\textbf{location\_is\_city\_center} &
Is the accommodation located in the city center\\
		\bottomrule
	\end{tabular}
\end{table*}

\begin{table*}[!h]
	\centering
	\caption{Training dataset field statistics}
	\label{tab:stat}
\begin{tabular}{lccccc}

\toprule
\multicolumn{1}{c}{\textbf{Field name}}                                  & \textbf{Unique Values} & \textbf{Mean Value} & \textbf{Mode Value} & \textbf{Min Value} & \textbf{Max Value} \\ 
\midrule

\textbf{review\_score}& 22                  & 8.87                   & 9.0               & 1.0                  & 10.0                  \\
\textbf{review\_helpful\_votes}& 50                      & 0.12                   & 0             & 0                  & 91                  \\
\textbf{guest\_type}& 4                   & Couple                   & -                & -                  & -                  \\
\textbf{guest\_country}& 271                      & -                   & Cobra Island		              & -                  & -                  \\
\textbf{room\_nights}& 72                    & 2.58                   & 4        & 1                  & 112                  \\
\textbf{month}& 12                    & -                   & July        & -                  & -                  \\
\textbf{accommodation\_type}& 27                    & -                   & Hotel        & -                  & -                  \\
\textbf{accommodation\_score}& 71                    & 8.38                   & 8.5        & 2.4                  & 10.0                  \\
\textbf{accommodation\_country}& 180                    & -                   & Australia        & -                  & -                  \\
\textbf{accommodation\_star\_rating}& 10                    & 2.38                   & 4.0        & 0.0                  & 5.0                  \\
\textbf{location\_is\_beach}& 2                    & 0.23                   & 0        & 0                  & 1                  \\
\textbf{location\_is\_ski}& 2                    & 0.02                   & 0        & 0                  & 1                  \\
\textbf{location\_is\_city\_center}& 2                 & 0.29                   & 0          & 0                  & 1                  \\
\bottomrule

\end{tabular}
\end{table*}

\subsection{Data Split}
The data is split to train, validation and test sets using 80\% / 10\% / 10\% random splits based on accommodation id. This means that every accommodation and its corresponding reviews will appear only in one of the sets. The dataset consists of the following files:
\begin{itemize}
    \item  \textbf{train.csv} - Training dataset of 1,628,989 reviews from 40,000 accommodations.
    \item  \textbf{validation.csv} - Validation dataset of 203,787 reviews from 5,000 accommodations.
    \item \textbf{test.csv} - Test dataset of 199,138 reviews from 5,000 accommodations.
\end{itemize}

Our repository currently consists only the training set (i.e., train.csv file) since we plan publishing a challenge in which the validation and test sets will not be exposed to its participants. These sets will be published right after the challenge.

\section{Problem Formulation} \label{problem_formulation}
Our objective is to create a model that predicts the helpfulness of every review tailored to individual users. In essence, we aim to construct a personalized helpfulness function, denoted as $f(r_j|c_i)$, which evaluates the relevance of review $r_j$ to user $i$ given their context $c_i$. This function assigns a score indicating the degree to which review $j$ is beneficial for user $i$. These scores enable us to rank reviews, ensuring that those with the highest $f$ values are deemed most helpful within context $c_i$.

Using the number of helpful votes as the target signal inherits multiple issues. First, it introduces a presentation bias towards the previous review ranking algorithm (usually sorted by votes). Additionally, the signal of votes is sparse as most of the reviews are not presented and therefore not voted. Moreover, there might be a cold-start problem where new reviews don't have as many votes as older reviews which might be less relevant over time. Finally, in many cases, only the final number of votes is stored and therefore it's not feasible to use this signal for developing personalized review ranking models.

Thus, we introduce a more feasible and novel approach for modeling personalized helpfulness measure. We propose to model the personalized helpfulness of a review as the likelihood that it is written by its reviewer given the reviewer's context. Notably, we define $f$ such that given a user context $c_i$, it estimates the likelihood that review $r_j$ was written by the user. Formally, we optimize $f$ such that given that review $r_i$ was written by a user with context $c_i$, it holds: 
\begin{equation}
\label{gt}
f(r_j|c_i) = \left\{
\begin{array}{ll}
1 & \text{if } i = j \\
0 & \text{if } i \neq j
\end{array}
\right.
\end{equation}

In practice, we learn $f$ through a contrastive modeling approach, described in Section~\ref{sec:method}.

\section{Method}
\label{sec:method}
\begin{figure*}[ht]
    \centering
\includegraphics[width=1\textwidth,height=0.5\textwidth]{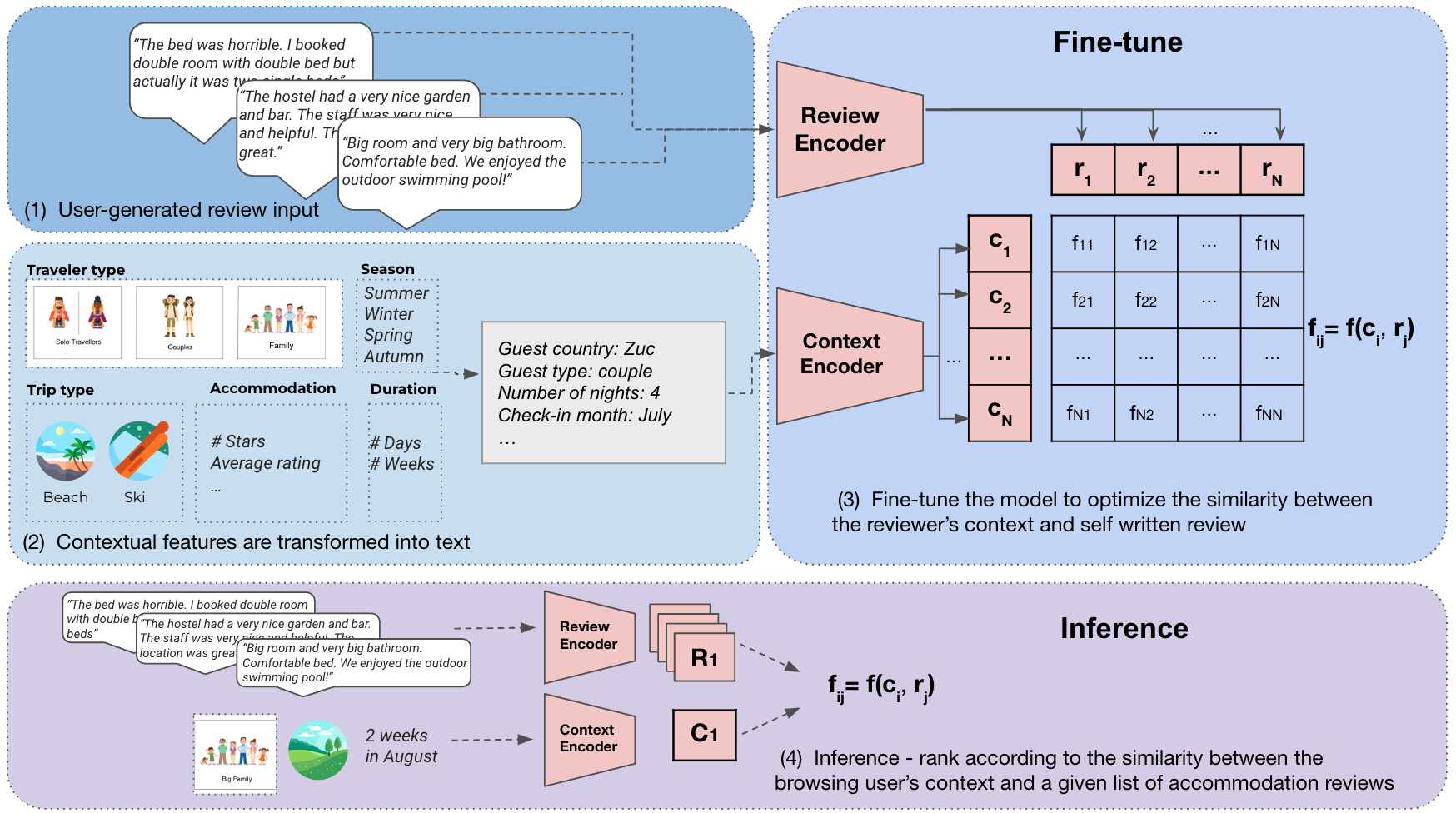}
        \caption{The proposed approach: (1) user-generated review and (2) user context are transformed into text and passed to (3) encoding layers that are fine-tuned to optimize $f_{i,j}$ diagonal to 1 and the rest to 0. (4) In inference time our model generates $f_{i,j}$ scores for a user context and a set of reviews, and ranks based on the descending order of scores.}
    \label{fig:method_diagram}
\end{figure*}

We propose a CL approach inspired by \citet{radford2021learning}. Our suggested approach entails several key steps. Initially, it consolidates the fields of the review into a unified string. Subsequently, it combines the fields representing the context into another string. Thirdly, both strings undergo separate encoding processes: one generates a latent representation for the context, while the other produces a latent representation for the review. Ultimately, our training objective aims to maximize the similarity between the latent representation of each review and its corresponding reviewer's context, while simultaneously minimizing the similarity between each review and contexts unrelated to the reviewer. In a live setting, our intention is to employ the resulting model to rank a list of accommodation reviews based on the context of the browsing user. Figure \ref{fig:method_diagram} offers a visual depiction of our approach. The subsequent paragraphs delve into a comprehensive explanation of each of the aforementioned steps.

\subsection{Encoding}
Both context and review can be described as a set of multiple fields of different types (textual, numeric and boolean). Inspired by \cite{hegselmann2023tabllm}, we consolidate all the review related fields into a single string and all the context related fields into another string.

We utilize two distinct encoders to process the context and review strings. Each encoder generates latent representations for $c_i$ and $r_j$ respectively. The encoder architecture incorporates a pre-trained language model designed to extract contextual embeddings from the input tokens. We adopt the $[CLS]$ token embedding as the latent representation for both the context and review strings.

\subsection{Interaction Matrix}
Given a batch of $N$ user context embeddings and their corresponding review embeddings, denoted as $C={c_1, ..., c_N}$ and $R={r_1, ..., r_N}$ respectively, where each review $r_i$ was written by a user with context $c_i$, we construct an interaction matrix $F$. Each element $f_{i,j}$ of $F$ signifies the similarity between context $c_i$ and review $r_j$. Aligning with the problem formulation described in Section ~\ref{problem_formulation}, we use a similarity function that guarantees $f_{i,j} \in [0, 1]$. Specifically, we use sigmoid function, i.e., $\sigma(x) = \frac{1}{1 + e^{-x}}$, over the dot product of the context and review embedding vectors:
\begin{equation}
f_{i, j} = \sigma({\mathbf{c_i} \cdot \mathbf{r_j}})
\end{equation}
\subsection{Loss Functions}
To optimize $F$ with respect to formula \ref{gt}, we experiment with the following loss functions:

\textbf{InfoNCE loss} - inspired by CLIP model loss (\citet{radford2021learning}),  we use n-pair / InfoNCE (Noise-Contrastive Estimation) loss, first introduced by \citet{sohn2016improved}. This loss maximizes the mutual information between positive pairs and minimizes it for negative pairs. We apply it on both row-wise and column-wise, and then take the average, as shown in the below formula: 
\begin{equation}
\mathcal{L}_{\text{InfoNCE}} = -\frac{1}{2N} \left( \sum_{i=1}^{N} \log \frac{\exp(f_{i,i})}{\sum_{j=1}^{N} \exp(f_{i,j})} + \sum_{j=1}^{N} \log \frac{\exp(f_{j,j})}{\sum_{i=1}^{N} \exp(f_{i,j})} \right)
\end{equation}
\textbf{Binary cross entropy (BCE) loss} - this loss computes binary cross entropy between the interaction matrix values and the targets, i.e., the identity matrix $I$: 
\begin{equation}
\mathcal{L}_{\text{BCE}} = -\frac{1}{N^2} \sum_{i=1}^{N} \sum_{j=1}^{N} [\mathbf{I_{i,j}} \log(f_{i,j}) + (1 - \mathbf{I_{i,j}}) \log(1 - f_{i,j})]
\end{equation}

\subsection{In-accommodation Sampling}
Usually, in contrastive learning, training batches are randomly sampled from the training set \cite{radford2021learning}. The assumption is that every sample is sufficiently distant from the other samples in the batch in terms of its concept / meaning. In our case, reviews from different accommodations can be easily differentiated. For example, reviews of a beach resort versus reviews of a hotel in an urban area. Therefore, randomly sampling the batch might result in a model that distinguishes between accommodations and ignores the user context. In addition, in a live setting reviews are always ranked against reviews from the same accommodation. Thus, we suggest an in-group sampling on the accommodation level where each batch consists of reviews and contexts from the same accommodation. We evaluate the efficacy of this sampling approach by comparing it to random sampling in our experiments.

\section{Experiments}
\subsection{Data Preprocessing}
To produce the textual inputs for the encoder, we preprocess the review and context fields. We create a textual representation for every field using the template: \verb|"<field_name>: <field_value>\n"|. Fields with empty values are skipped. Then, we concatenate the textual representations of the fields according to a predefined order. The review input string consists of the following fields in the following order: review title, review positive, review negative and guest score. The context input string consolidates user-related fields and accommodation-related fields. The user related fields occur in the following order: guest country, guest type, number of nights and check-in month. The accommodation-related fields are organized in the following order: accommodation type, accommodation star rating, accommodation score, accommodation type, location is beach, location is ski and location is city center.

\subsection{Encoder Architecture}
We fine-tuned \verb|all-MiniLM-L6-v2| model from Hugging Face Sentence Transformers repository\footnote{\url{https://huggingface.co/sentence-transformers}}. This model is pre-trained on a variety of semantic similarity prediction tasks using a contrastive objective over above 1 billion sentence pairs. Considering other models and architectures might be beneficial as well, however, finding an optimal combination of such is not a central aim of this research.

\subsection{Fine-tune Details}
We applied AdamW optimizer \cite{loshchilov2018fixing} with a weight decay of \verb|0.01| and an initial learning rate of \verb|3e-5|. We fine-tuned for 4 epochs as observed to be enough for the fine-tune process to saturate. We employed a batch size of 64 and a warm-up rate of 0.05. We experimented with batch sizes of $[16, 32, 64, 128]$ and didn't observe significant differences in performance. All of our experiments were performed on a computation instance equipped with 1 NVIDIA A10G Tensor Core GPU, 8 vCPU and 32GB RAM. The fine-tune process took 9 hours.

\subsection{Baselines}
We compare our approach with two baselines: (1) helpful votes ranking - in which $f_{i,j}$ are assigned with the number of helpful votes $r_j$ has. Ranking is performed based on the descending order of $f_{i,j}$ values. (2) Pre-trained model - in which we used \verb|all-MiniLM-L6-v2| pre-trained model without fine-tuning it.

\begin{table*}
\centering
\caption{Performance of our approach and baselines. The highest results for each of the metrics are highlighted in bold font.}
\label{tab:results}
\begin{tabular}{lcccccc} 
\toprule
\textbf{}                          & \multicolumn{2}{c}{\textbf{MRR}} & \multicolumn{2}{c}{\textbf{Precision@1}} & \multicolumn{2}{c}{\textbf{Precision@10}} \\ 
\textbf{Model}                     & \textbf{Mean} & \textbf{Std} & \textbf{Mean} & \textbf{Std} & \textbf{Mean} & \textbf{Std} \\
\midrule
Helpful votes baseline                  & 0.096 & 0.039          & 0.025 & 0.034                & 0.248 & 0.097               \\
Pre-trained baseline                    & 0.098 & 0.039          & 0.026 & 0.035                & 0.251 & 0.095                \\
\hline
Random sampling, InfoNCE loss           & 0.147 & 0.051          & 0.049 & 0.048                & 0.375 & 0.106                \\
Random sampling, BCE loss               & 0.191 & 0.063          & 0.089 & 0.064                & 0.419 & 0.111                \\
In-accommodation sampling, InfoNCE loss & 0.237 & 0.067          & 0.111 & 0.069                & 0.519 & 0.111 \\
In-accommodation sampling, BCE loss     & \textbf{0.278} & 0.074 & \textbf{0.154} & 0.079       & \textbf{0.549} & 0.110                 \\
\bottomrule
\end{tabular}
\end{table*}

\subsection{Metrics}

We segment the test set into groups of context-review tuples from the same accommodation. We denote the set of accommodations with $A$ and the $k^{th}$ accommodation with $a_k$. For every accommodation $a_k$ there is a set of context-review tuples $C_k$ and $R_k$ such that $|C_k| = |R_k|$, and every review $r_i \in R_k$ was written by the reviewer with context $c_i \in C_k$. For every $c_i \in C_k$ we produce a ranked list of the reviews within the same accommodation $R_k$. We denote the rank of $r_j$ (the review written by the user with context $c_j$) in the ranked list of $c_j$ with $Rank(j)$. Note that our goal is that given a context $c_j$, review $r_j$ should have the highest rank in the ranked list of reviews.

We use 3 common metrics from the world of recommendation:
\textbf{Mean Reciprocal Rank (MRR)} - we formulate the MRR metric as follows:
\begin{equation}
MRR = \frac{1}{|A|} \sum_{k=1}^{|A|} \frac{1}{|C_k|} \sum_{j=1}^{|C_k|} \frac{1}{Rank(j)}
\end{equation}
\textbf{Precision@k} - motivated by that reviews are usually presented in pages of 10 reviews per page, we measure precision@k for $k \in [1, 10]$:
\begin{equation}
Precision@k = \frac{1}{|A|} \sum_{k=1}^{|A|} \frac{1}{|C_k|} \sum_{j=1}^{|C_k|} {\text{I}(Rank(j) \leq k)}
\end{equation}

\subsection{Results}

\begin{figure}[h]
    \centering
\includegraphics[width=0.475\textwidth,height=0.24\textwidth]{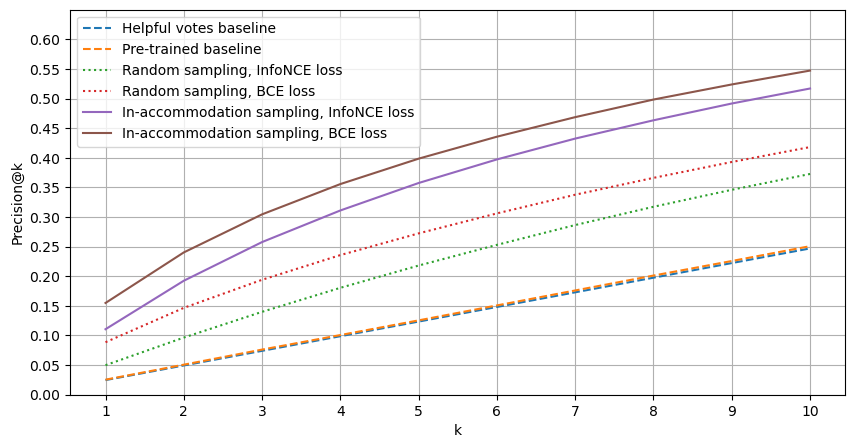}
        \caption{Precision@k over $k \in [1,10]$}
    \label{fig:precisionk}
\end{figure}

Results are described in Table~\ref{tab:results} and Figure~\ref{fig:precisionk}. We performed a Friedman test \cite{milton1939correction} that showed there was a statistically significant difference between the methods (p-value<0.001). A post-hoc analysis using the pairwise Dunn test \cite{dinno2015nonparametric} showed that the model combines in-accommodation sampling with BCE loss significantly outperformed all other methods (p-value<0.001). Moreover, the in-accommodation sampling significantly outperformed random sampling for both InfoNCE and BCE loss functions (p-value<0.001). Finally, the BCE loss achieved significantly better performance compared to InfoNCE loss (p-value<0.001). This is probably due to the softmax applied in the InfoNCE loss which normalizes the similarity values one time on the row level and another time on the column level. The BCE, by not applying softmax, allows multiple $f_{i, j}$ to have high values (i.e., close to 1) on the same row / column without forcing any dependency between them.

\begin{table*}
\centering
	\caption{Comparative analysis. 'Original Review Text' is the original text written by the reviewer, 'Best Performing Model’s Top Review' is our model top ranked review (excluding the original review), 'Pre-trained Baseline Model’s Top Review' is the baseline top ranked review (excluding the original review). It can be clearly noticed that our model shares much more common topics with the original review compared to the baseline, showing our model better captures the user segment needs.}
	\label{tab:examples}
\begin{tabular}{p{0.15\textwidth}p{0.25\textwidth}p{0.25\textwidth}p{0.25\textwidth}}
\toprule
\textbf{Guest Type}  & \textbf{Original Review Text}                                                                                                                                                                                                                                  & \textbf{Best Performing Model's Top Review}                                                                                                                                                                                                                                   & \textbf{Pre-trained Baseline Model's Top Review}                                                                                                                                                                                  \\
\midrule
couple               & \textcolor[rgb]{0.6,0,1}{Fantastic views} across the Sassi from \textcolor[rgb]{0.576,0.769,0.49}{terrace}. \textcolor{red}{Fabulous location}. Easy access to property. \textcolor[rgb]{1,0.6,0}{Helpful and communicative host}.                             & \textcolor{red}{Fantastic location} with \textcolor[rgb]{0.6,0,1}{excellent view}. Comfortable and well laid out small apartment. \textcolor[rgb]{1,0.6,0}{Very responsive and helpful host}.                                                                                      & Fantastic place! Wonderful \textcolor[rgb]{0.714,0.843,0.659}{terrace} and very comfortable and well equipped place.                                                                                                     \\
\hline
couple               & \textcolor[rgb]{0.6,0,1}{The view} and the pool are amazing. \textcolor[rgb]{1,0.6,0}{The owner try is best to make your stay good}. \textcolor[rgb]{0.29,0.525,0.91}{Not to far from} the shops and \textcolor[rgb]{0.29,0.525,0.91}{beaches}.                & - the location \textcolor[rgb]{0.29,0.525,0.91}{near the best beaches}\par{}- \textcolor[rgb]{0.6,0,1}{the view} from the rooms and from the pool is insane \par{}- \textcolor[rgb]{1,0.6,0}{the kindness of the family owner}.                                                    & Perfect stay. \textcolor[rgb]{0.6,0,1}{Great} hammock and \textcolor[rgb]{0.6,0,1}{views}. My room was quite a ways up the hill and I liked it and the view from there. It was peaceful.                                 \\
\hline
family with children & Very \textcolor{red}{close to Retiro Park}, about 1.5 kms from the normal touristy areas. Quiet surroundings. \textcolor[rgb]{0.6,0.6,0.6}{Huge room} 305- immaculate bathroom.                                                                                & Superb location, \textcolor{red}{next to Retiro} and Prado Museum. Very convenient facilities, with \textcolor[rgb]{0.718,0.718,0.718}{family rooms}, and easy check in process. Staff was attentive.                                                                              & Located in my favorite spot in Madrid, cool Lobby area and good breakfast to start off the day.                                                                                                                          \\
\hline
family with children & The receptionist was welcoming and the room was spacious and it has \textcolor[rgb]{0.706,0.655,0.839}{amenities like shops} with free delivery and also a \textcolor[rgb]{0.902,0.722,0.686}{pool}.                                                           & Like the \textcolor[rgb]{0.918,0.6,0.6}{pool} \& the playground kids really enjoyed also the place is accessible a lot of \textcolor[rgb]{0.706,0.655,0.839}{options were you can buy some stuff}.                                                                                 & Very good value large apartment. Large \textcolor[rgb]{0.902,0.722,0.686}{pool} outside, good location.                                                                                                                  \\
\hline
solo traveller       & \textcolor[rgb]{1,0.6,0}{Very attentive and friendly staff!} Everything was clean and the other guests were nice, huge rooftop to sit in the evening and enjoy the fresh breeze.                                                                               & I like the inner patio. The location is great! \textcolor[rgb]{1,0.6,0}{I liked the owner and the manager} at the hotel. I like there is the washing machine and dryer.                                                                                                            & Best location, close to many main places, nice room.                                                                                                                                                                     \\
\hline
solo traveller       & \textcolor[rgb]{0.6,0,1}{Beautiful setting} and great value for money. \textcolor[rgb]{1,0.851,0.4}{Location is safe} and easy access to the highway. Loved the shower with a view of the hill and the \textcolor[rgb]{1,0.6,0}{staff is very friendly}.       & The \textcolor[rgb]{0.6,0,1}{place was beautiful} and in a good neighbourhood that \textcolor[rgb]{1,0.851,0.4}{felt safe}. The host and \textcolor[rgb]{1,0.6,0}{staff were very friendly}. Stayed only one night and hopefully I will stay longer in the future.                 & Personal touch. \textcolor[rgb]{1,0.6,0}{Gina was a lovely host}. \textcolor[rgb]{0.6,0,1}{Beautiful views} over Johannesburg.                                                                                           \\
\hline
group                & \textcolor[rgb]{0.8,0.255,0.145}{Great place.} The beds were really comfortable and \textcolor[rgb]{0.463,0.647,0.686}{the place is quiet}. They \textcolor[rgb]{0.867,0.494,0.42}{offer tours} to the Aral Sea, which is why we booked here. Would recommend. & Very helpful staff, great breakfast. \textcolor[rgb]{0.463,0.647,0.686}{Nice relaxed yard, calm} and clean rooms. Thank you also for helpful \textcolor[rgb]{0.867,0.494,0.42}{organization of tour} to Elliq Qala.                                                                & The hostel is a very \textcolor[rgb]{0.8,0.255,0.145}{nice place}. We had a large room and in the morning we took off for Muynak, so we didn't have breakfast at the place. People at the reception desk are very kind.  \\
\hline
group                & I like the breakfast. \textcolor{red}{Location is good}, very \textcolor[rgb]{0.835,0.651,0.741}{close to the old town center}. Room is not very big, but \textcolor[rgb]{0.718,0.718,0.718}{comfortable}.                                                     & \textcolor{red}{Location is premium}, with \textcolor[rgb]{0.835,0.651,0.741}{easy access} to all city areas (\textcolor[rgb]{0.835,0.651,0.741}{especially Old Town}). \textcolor[rgb]{0.718,0.718,0.718}{The room is very comfortable}, and the common areas are well decorated. & Spotless modern hotel in an \textcolor{red}{excellent area} close to the main Sq. \\

\bottomrule
\end{tabular}
\end{table*}

\subsection{Comparative Analysis}

In the following section, we delve into a comparative analysis of the results generated by our best performing model against the pre-trained baseline. Our aim is to demonstrate the interpretability and explainability of our model, thereby improving transparency and fostering trust.

Table~\ref{tab:examples} presents 8 comparisons of the best performing model (i.e., in-accommodation sampling with BCE loss) vs the pre-trained baseline model. For simplicity, we show only part of the guest's context (i.e., the guest type) and only the positive ("liked") section of the reviews. However, the process we perform can be applied over the full set of fields. The table describes the original review (written by the guest), the model's top review and the pre-trained baseline's top review. In case any of the models selected the original review to be on top of the ranked list, we show the second result in the list. We do this to simulate a production setting where the user hasn't written the review yet. We selected reviews with 100-200 characters for ease of comparison. Examples were randomly selected, 2 for every guest type. We identified the topics mentioned in the reviews using the Text2topic model \cite{wang2023text2topic} and colored the common topics between the original reviews and the models' top reviews. For example, the first row in the table describes a review written by a couple. The review highlights the views, the terrace, the location, and the host helpfulness. While the pre-trained baseline selected a review that highlights the terrace, our best performing model selected a review that highlights the location, the views and the host helpfulness but without mentioning the terrace.

As can be seen, this analysis provides an insightful visual comparison between models. It enables to identify which model has a higher topic intersection with the guest original review which indicates how effectively it interprets contexts into user preferences. Moreover, it enables to identify the topic intersection between the models, and whether different models identifies different topics. Such insight for example, might motivate model ensemble. Finally, it enables us to learn and explore the preferences of the different user segments.

\section{Conclusions and Future Work}
In this paper, we introduce a comprehensive review dataset sourced from a prominent online travel platform. Our work proposes a novel formulation for personalized review ranking, aiming to mitigate common biases and challenges observed in current methodologies. We employ a contrastive learning approach to tackle this task and evaluate its efficacy across various experimental setups. Through comparative analysis, we demonstrate how our end-to-end solution effectively captures the interplay between a reviewer's context and their review, thus enabling a personalized review ranking experience.

Our future work includes deploying and conducting experiments with our best performing model in production via an A/B test. We seek to quantify user engagement metrics and assess the business impact of our approach. Furthermore, we strive to publish a challenge encouraging research individuals and groups to address the task based on our dataset and problem formulation. Additionally, we plan to enrich both context and review data by integrating new signals such as user-provided review images. Finally, we aim to expand our dataset to include more languages, enabling the development of multilingual models to better serve diverse user segments.

\bibliographystyle{ACM-Reference-Format}
\bibliography{custom}

\appendix

\end{document}